\documentclass[aps,prl,twocolumn,preprintnumbers,amsmath,amssymb]{revtex4}

\usepackage{graphicx}% Include figure files
\usepackage{bm}% bold math

\begin{document}

\title{Addressing a single NV$^{-}$ spin with a macroscopic dielectric microwave cavity}

\author{J.-M. Le Floch\textsuperscript{1,2}}
\author{C. Bradac\textsuperscript{2,3}}
\author{N. Nand\textsuperscript{2,3}}
\author{S. Castelletto\textsuperscript{4}}
\author{M.E. Tobar\textsuperscript{1,2}}
\author{T. Volz\textsuperscript{2,3}}
\affiliation{\textsuperscript{1}School of Physics, The University
of Western Australia, Crawley, WA 6009, Australia \\
\textsuperscript{2}ARC Centre of Excellence for Engineered Quantum
Systems \\ \textsuperscript{3}Department of Physics and Astronomy,
Macquarie University, North Ryde, NSW 2109, Australia \\
\textsuperscript{4}School of Aerospace, Mechanical and Manufacturing
Engineering, RMIT University, Melbourne, Australia }
\date{\today}

%\pacs{xxx}
\begin{abstract} We present a technique for addressing single
NV$^{-}$ center spins in diamond over macroscopic distances using a
tunable dielectric microwave cavity. We demonstrate optically
detected magnetic resonance (ODMR) for a single NV$^{-}$ center in a
nanodiamond (ND) located directly under the macroscopic microwave
cavity. By moving the cavity relative to the ND, we record the ODMR
signal as a function of position, mapping out the distribution of
the cavity magnetic field along one axis. In addition, we argue that
our system could be used to determine the orientation of the
NV$^{-}$ major axis in a straightforward manner.
\end{abstract}

\maketitle

\begin{figure}[ht!]
  \centering
     \includegraphics[width=86mm]{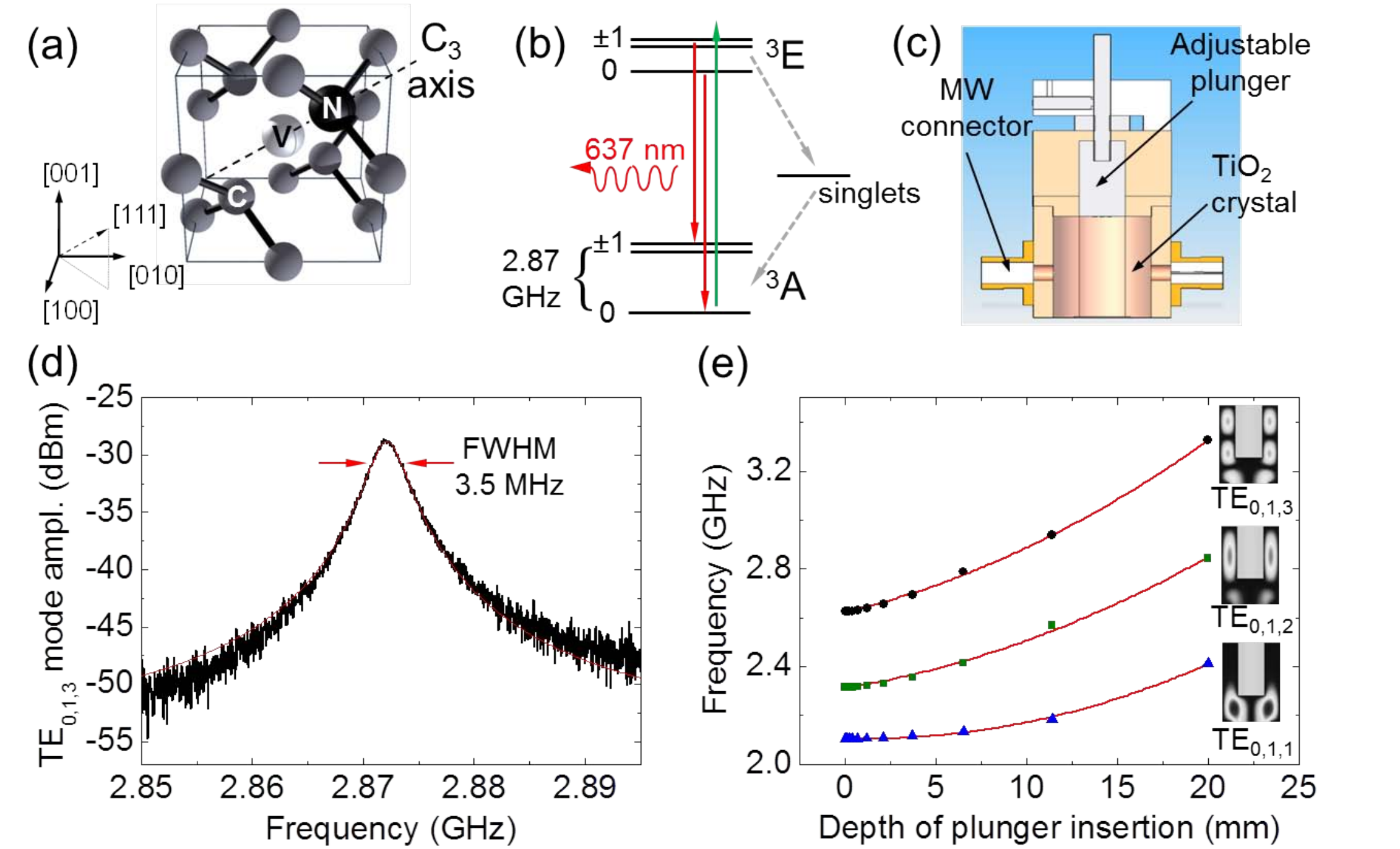}
  \caption{(a) Diamond lattice structure with an embedded NV center.
  (b) Level scheme for the NV$^{-}$ center including the hyperfine splitting
  of the triplet ground state ($^3$A). Optical pumping of the spin into the
  $m_S=0$ state occurs via an intersystem crossing to the singlet manifold.
  (c) Dielectric cavity with adjustable plunger. The outer diameter of the cavity is
  32~mm, its height amounts to 20~mm. (d) Transmission spectrum of
  the cavity with a FWHM of 3.5~MHz corresponding to a Q-factor of around
  1000. The spectrum was recorded using a Fieldfox N9918A (Agilent
  Technologies). (e) Numerically calculated cavity frequency as a function of plunger position.
  The mode of interest TE$_{0,1,3}$ tunes easily across the NV$^{-}$ ground-state spin transition.
  \label{Fig1}}
\end{figure}

Over the past decade, nitrogen-vacancy (NV) color centers in diamond
(Figure~\ref{Fig1}(a)) have attracted a great deal of interest due
to their outstanding quantum properties \cite{Doherty:PRep13}.
Experiments have demonstrated long ground-state spin coherence times
of the negatively charged NV center (NV$^{-}$) even at room
temperature \cite{Balasubramanian:NatMat09}. This makes NV$^{-}$
centers in diamond ideal candidates for room-temperature qubits
\cite{Dutt:Science07,Neumann:Science08} and for ultrasensitive
quantum sensors for detecting magnetic
\cite{Balasubramanian:Nature08,Maze:Nature08,Degen:APL08,Taylor:NatPhys08,McGuinness:NatNano11,Grinolds:NatNano14,Tetienne:Science14}
and electric fields \cite{Dolde:NatPhys11} at the nanoscale even in
biological settings
\cite{Faklaris:Small08,McGuinness:NatNano11,Ermakova:NanoLett13}.

Both quantum information processing and quantum sensing with
NV$^{-}$ spins require the (coherent) manipulation and addressing of
individual spins typically through the application of microwave (MW)
radiation at a frequency that is resonant with the ground-state spin
transition. The NV$^{-}$ exhibits a zero-field spin resonance at
2.87 GHz, which occurs between the $m_s = 0$ and $m_s = \pm 1$ spin
sublevels of its spin triplet ground state (see Figure 1b). The most
commonly used approaches for applying microwaves at this frequency
are on-chip microstrip lines (thin wires), coplanar waveguides, or
free-space loop antennas. On-chip solutions rely on near-field
coupling and require the NV$^{-}$ spin to be in close proximity (on
the order of $10~\mu$m) to the wire or microstrip line. Besides the
inhomogeneity of the field, these on-chip solutions can easily lead
to significant sample heating and undesired sample drift. Loop
antennas typically work in the far field but require orders of
magnitude larger amount of radiated MW power.

In order to address single NV$^{-}$ spins in diamond, we designed a
so-called dielectric-loaded microwave resonator (DLR) with high
quality (Q) factor (see Figure~\ref{Fig1}(c)). DLRs of this kind are
typically used in low-temperature electron paramagnetic resonance
experiments for measuring the complex permittivity of extremely
low-loss dielectrics~\cite{Tobar:JAP98, Krupka:MST99,
LeFloch:PLA06}, but are also employed for testing local Lorentz
invariance in fundamental experiments \cite{Tobar:PRD09}. In
industrial settings, DLRs find applications in radar, proximity
detection, information transmission, remote guiding, navigation and
positioning \cite{Rieck:Proc08, Droz:Proc06}. In order to find an
appropriate design for our DLR, we employed the numerical method of
lines \cite{LeFloch:IEEE07} or finite element analysis. The design
was guided by the desire to have compact cavity dimensions and the
requirement for the field to couple evanescently to the NV$^{-}$
spins located in close vicinity under the cavity.

We found the best configuration to be an open cylindrically
symmetric cavity with a pure transverse electric (TE) mode with two
non-vanishing magnetic-field components, H$_r$ and H$_z$. In
contrast to whispering gallery modes, the TE-field confinement into
the dielectric is not as high and exhibits less spurious modes,
leading to a higher Q-factor \cite{LeFloch:RevSciInst14}. We denote
the different cavity modes by TE$_{m,n,p}$, where $2m$ is the number
of azimuthal nodes, $n$ the number of radial nodes, and $p$ the
number of nodes along the z-axis (symmetry axis) of the cylinder.
For pure TE-modes the azimuthal mode number vanishes, i.e. $m = 0$.
Due to the particular boundary conditions, they only have three
non-vanishing components of the electromagnetic field, $E_{\Theta},
H_{r}$ and $H_{z}$. By inserting a dielectric rod made of
high-permittivity, low-loss microwave material, the field can be
confined to an area of roughly $10 \times 10$~mm$^2$. From the few
suitable materials available~\cite{LeFloch:APL08}, we chose TiO$_2$
for which the fundamental mode has a frequency of 2.2~GHz. For
addressing the NV$^{-}$ spin transition, we then use the
higher-order TE$_{0,1,3}$-mode resonating at 2.7~GHz. Frequency
tuning is achieved by mechanical insertion of a metallic plunger
which directly affects the electric field and shifts the resonance
frequency up to a value of 3.1~GHz (Figure~\ref{Fig1}(e)). The
(loaded) Q-factor of this cavity mode is about 1,000 (see
Figure~\ref{Fig1}(d)). Note that for perfect input coupling we would
expect the circulating intra-cavity power to be enhanced by a factor
Q compared to the incoming MW power. However, in the current cavity
the coupling coefficient is rather small (about 1$\%$).

\begin{figure}[ht!]
  \centering
     \includegraphics[width=86mm]{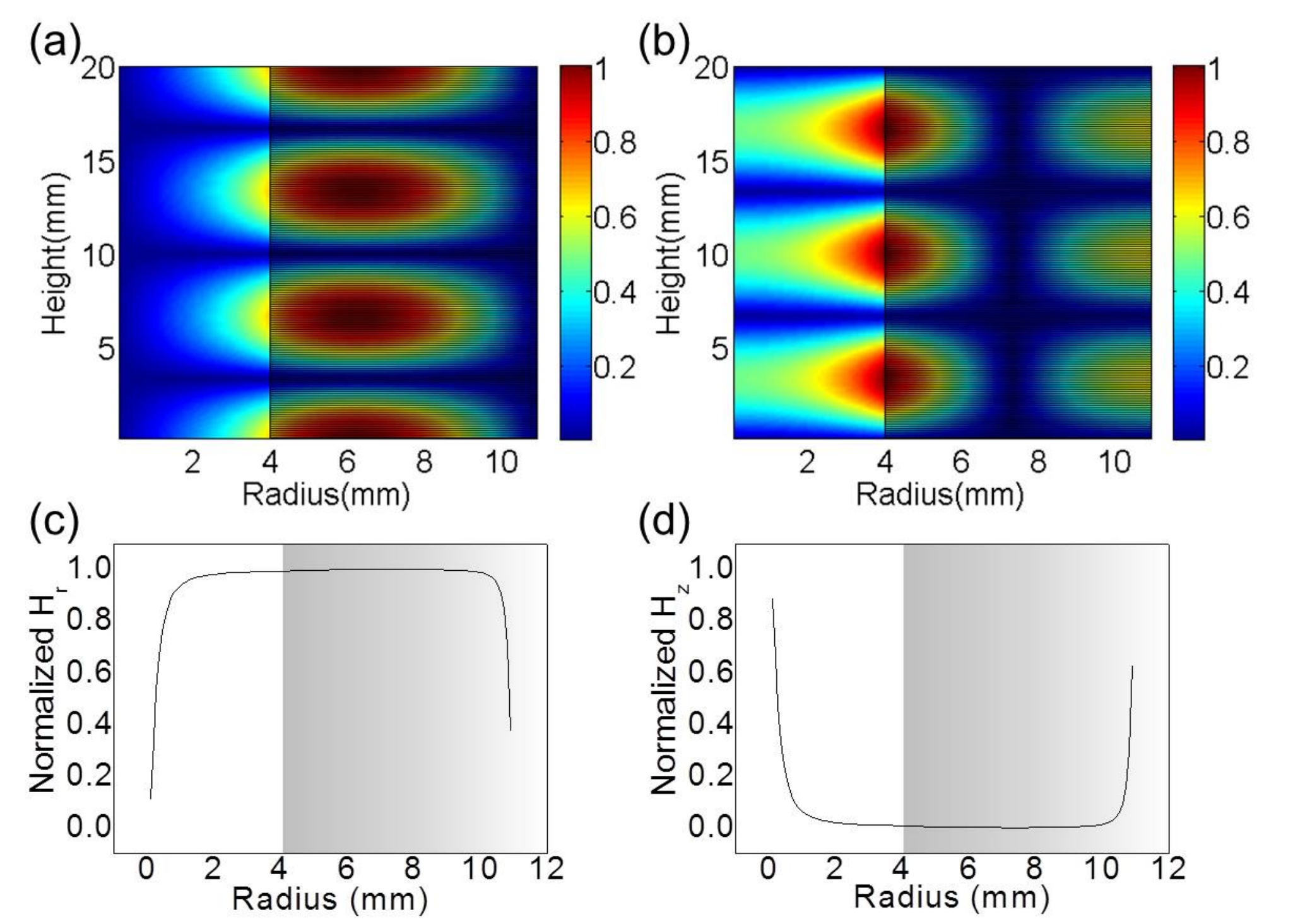}
  \caption{(a) Radial magnetic field strength $H_r$ in a two-dimensional cut
  along the vertical symmetry axis (y-axis of the graph) of the microwave cavity.
  (b) Analogous plot for the strength of the vertical magnetic field $H_z$. Shaded
  areas indicate the dielectric material of the cavity.
  (c) and (d) Field intensities $H_r$ and  $H_z$ along the radial direction in a plane $1$~mm below the cavity.
  Both fields are normalized to the local overall field strength $\sqrt{H^2_r + H^2_z}$.
  In the center of the trap, only $H_z$ is non-vanishing whereas directly under the dielectric
  (shaded region) $H_r$ is the dominant field component.
  \label{Fig2}}
\end{figure}

%The coupling factors ($\beta_{in}$ and $\beta_{out}$) between the
%input/output transmission lines (more specifically coaxial loop
%probes) and the cavity are needed to obtain the unloaded $Q$-value,
%$Q_0$, from the measured loaded Q value, $Q_L$, using the following
%expression: \begin{equation}   Q_0 = Q_L\left(1+ \beta_{in} +
%\beta_{out}\right)
%\end{equation}
%The coupling factors can be obtained by measuring the reflected and
%transmitted power \cite{Zamarro:Tech87}. For small enough coupling,
%one can approximate $\beta = \beta_{in} = \beta_{out}$ and obtain
%$\beta$ simply from the transmitted power. In our cavity the
%coupling coefficient is rather small, $\beta \approx 10^{-2}$. In
%principle, the coupling coefficient can be engineered very close to
%unity without affecting the unloaded Q factor of the cavity. The
%power circulating in the dielectric of the cavity is given by
%\begin{equation}
%  P_{circ} = P_{in} \frac{Q_L}{2\pi} \frac{4 \beta_{in}}{\left(1 + \beta_{in} + \beta_{out} \right)^2}
%\end{equation}
%Hence, we see that unlike in a normal ODMR setup using a simple loop
%antenna, the circulating power in the cavity is enhanced compared to
%the power coupled into the cavity by the (loaded) Q factor of the
%cavity. In our case this would correspond to an enhancement by three
%orders of magnitude in the ideal case $\beta=1$.

\begin{figure}[ht!]
  \centering
     \includegraphics[width=86mm]{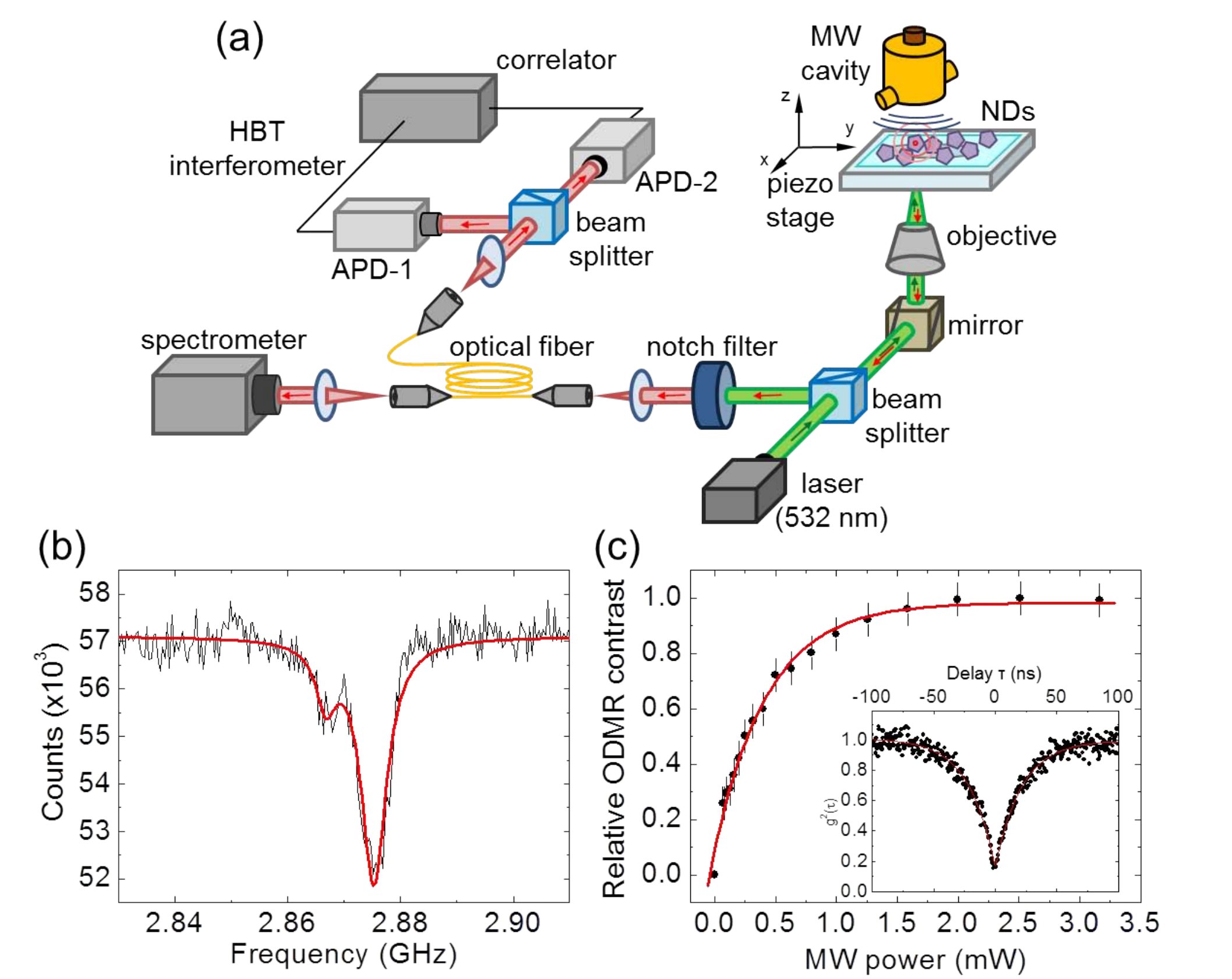}
  \caption{(a) Experimental setup for measuring ODMR with the dielectric MW
  cavity. The ND fluorescence is collected from below using a confocal
  setup. The collected photons are sent to either a spectrometer or
  to avalanche photodetectors.
  (b) Single-spin ODMR signal for a ND in the dielectric MW cavity.
  (c) Saturation curve for the ODMR contrast as a function of microwave power (after the signal generator). The inset shows
  an autocorrelation signal with clear antibunching demonstrating that the ND contains a single NV$^{-}$ center.} \label{Fig3}
\end{figure}

The magnetic field of the DLR cavity has cylindrical symmetry. Both
the radial magnetic field strength, $H_r$, and the vertical magnetic
field strength, $H_z$, are displayed in Figure~\ref{Fig2}.
Figures~\ref{Fig2} (a) and (b) show two-dimensional plots of the
respective field strength in a plane that contains the symmetry axis
of the cavity: The x-axis corresponds to the radial distance from
the symmetry axis and the values on the y-axis indicate the height
above the bottom edge of the cavity. We note that the radial field
component $H_r$ has a maximum right at the bottom edge at $r \approx
7$~mm, indicating a strong evanescent component. The vertical field
$H_z$ is well-contained within the cavity. Figure~\ref{Fig2} (c) and
(d) show the expected variation of the radial and vertical field in
a plane $1$~mm below the cavity. The plots are normalized to the
local overall field strength $\sqrt{H^2_r + H^2_z}$. The graphs
clearly show that in the center of the cavity the $H_z$ component
dominates (due to symmetry) whereas right under the dielectric at $r
\approx 7$~mm, the radial field is the dominant component.

We now move on to demonstrate ODMR of a single NV$^{-}$ spin located
just below the cavity. The HPHT nanodiamonds (MSY 0.1 $\mu$m,
Microdiamant) are placed on a glass coverslip approximately 1~mm
below the cavity which is mounted on a x-y-z mechanical stage (see
Figure~\ref{Fig3}(a)). The ND fluorescence upon excitation with a
532~nm laser is collected using a home-built confocal microscope
\cite{Bradac:NatNano10} and sent to either a spectrometer or to
avalanche photodetectors. Once a suitable single NV$^{-}$ center is
identified, we obtain an ODMR signal by applying microwave radiation
through our microwave cavity and recording the corresponding
fluorescence as a function of microwave frequency. The microwave
signal is generated using a standard microwave generator (SMIQ 06B,
Rohde $\&$ Schwarz) and amplified (25S1G4A, Amplifier Research)
before applying it to the cavity. A typical ODMR signal is displayed
in Figure~\ref{Fig3}(b), clearly demonstrating the coupling of a
single NV$^{-}$ spin to the macroscopic microwave resonator. Note
that the contrast of the ODMR signal was optimized by adjusting the
cavity resonance frequency to the actual transition frequency of the
selected NV$^{-}$ center. Depending on the ND, we found a maximum
achievable contrast of up to $12\%$. Next, we recorded a saturation
curve for the $m_s=0 \rightarrow m_s = \pm 1$
transition~\ref{Fig3}(c) giving a saturation power of about 5~dBm
for this particular NV$^{-}$ spin.

\begin{figure}[ht!]
  \centering
     \includegraphics[width=86mm]{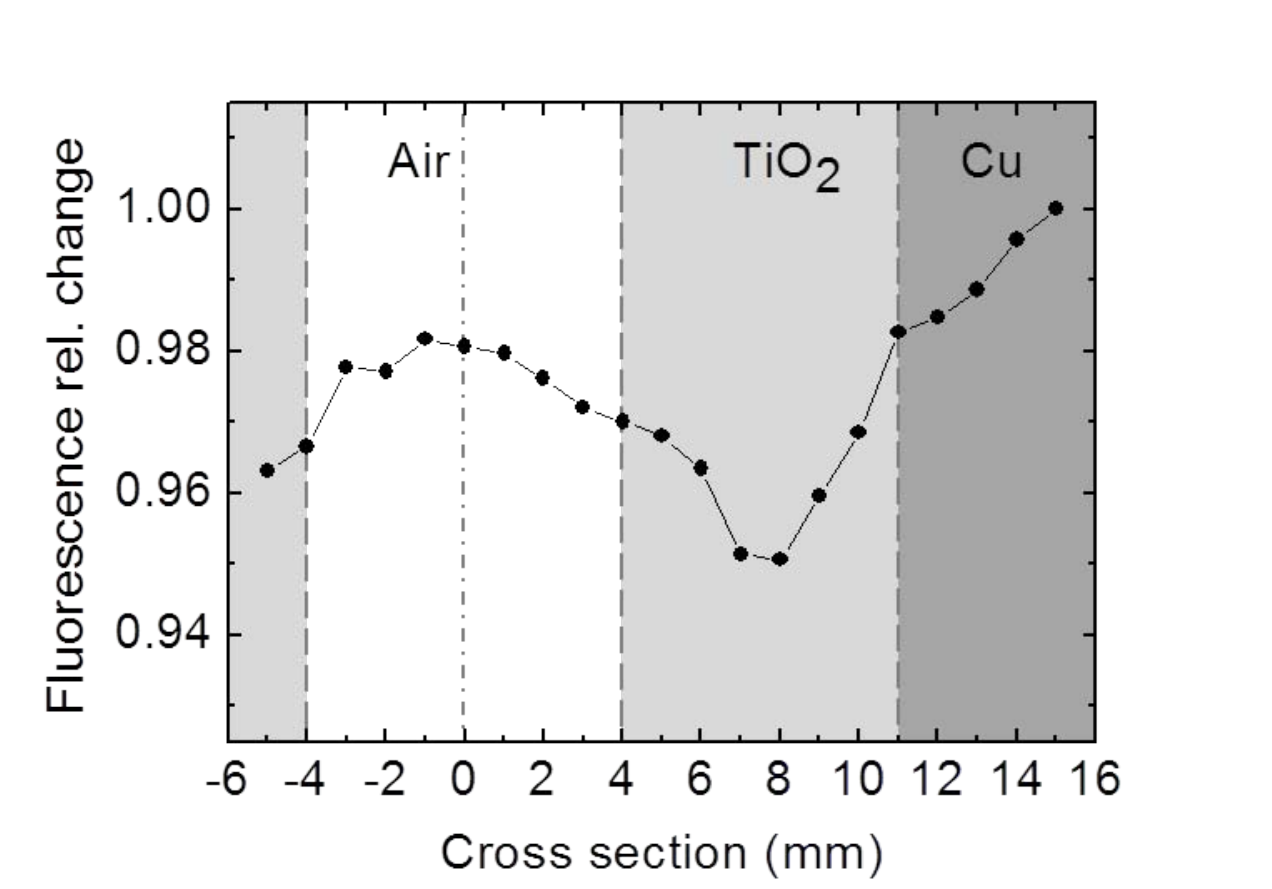}
  \caption{ODMR contrast as a function of ND position relative to the cavity axis.
  The maximum contrast is observed directly under the dielectric material in agreement
  with the expected field dependence from Figure~\ref{Fig2}. At the center of the
  cavity ($r=0$), the contrast does not completely vanish due to the finite value of $H_z$.
  } \label{Fig4}
\end{figure}

In order to demonstrate the spatial variation of the magnetic field,
we recorded an ODMR signal as a function of relative position
between the ND and the center axis of the cavity by mechanically
adjusting the cavity position. In the low-power (or linear) regime,
the contrast of the ODMR signal measures the local microwave power
seen by the NV$^{-}$ spin. Figure~\ref{Fig4} shows the result of
such a measurement taken along the x-axis in consecutive 1-mm steps,
while the z- and y-coordinates of the resonator were kept fixed with
respect to the ND position. In the figure, we plot the depth of the
ODMR resonance as a function of position, normalized to 1. The plot
displays the expected variation in ODMR contrast and exhibits a
maximum contrast of about 6$\%$ when the ND is right below the
dielectric of the cavity structure at $r \approx 7$~mm. The finite
contrast at $r=0$ indicates that the NV$^{-}$ spin has a
non-vanishing in-plane component. Since we do not know the major
axis of the NV$^{-}$ center, we cannot extract the full information
about the cavity magnetic field from Figure~\ref{Fig4}. Using a ND
with a known spin orientation, however, magnetic-field tomography is
possible. The cavity presented here could also be used as a tool to
determine the major axis of a single NV$^{-}$ center in a
straightforward manner: By measuring the ODMR contrast in the center
of the cavity and in two additional points along the circumference
just below the dielectric slab at $r \approx 7$~mm, the direction of
the NV$^{-}$ spin can be calculated - provided the ratio
$H_z(r=0)/H_r(r=7\,\mathrm{mm})$ is known.

The main advantages of our new technique are the large area over
which the spins can be addressed and the absence of undesired sample
heating allowing for stable long-term observations. Our measurements
once again demonstrate the potential of NV$^{-}$ centers as robust
technologically viable magnetic-field sensors. In addition, the DLR
cavity can serve as a tool for identifying the orientation of
NV$^{-}$ spins. In the future, we plan to use the device for
coherent time-resolved spin manipulation.


\begin{thebibliography}{99}

%1
\bibitem{Doherty:PRep13}
M. W. Doherty,N. B. Manson, P. Delaney, F. Jelezko, J. Wrachtrup,
and L. C. L. Hollenberg,
%The nitrogen-vacancy colour centre in diamond,
\textit{Physics Reports} \textbf{528}, 1 (2013).

%2
\bibitem{Balasubramanian:NatMat09}
G. Balasubramanian \textit{et al.},
%Ultralong spin coherence time in isotopically engineered diamond,
\textit{Nature Materials}, \textbf{8}, 383-387 (2009).

%3
\bibitem{Dutt:Science07}
%M.V.G. Dutt \textit{et al.},
M. V. Gurudev Dutt1, L. Childress, L. Jiang, E. Togan, J. Maze, F.
Jelezko, A. S. Zibrov, P. R. Hemmer, and M. D. Lukin
%Quantum Register Based on Individual
%Electronic and Nuclear Spin Qubits in Diamond.
\textit{Science}, \textbf{316}, 1312-1316 (2007).

%4
\bibitem{Neumann:Science08}
%P. Neumann \textit{et al.},
P. Neumann, N. Mizuochi, F. Rempp, P. Hemmer, H. Watanabe, S.
Yamasaki, V. Jacques, T. Gaebel, F. Jelezko, and J. Wrachtrup
%Multipartite Entanglement Among Single Spins in Diamond.
\textit{Science} \textbf{320}, 1326-1329 (2008).

%5
\bibitem{Balasubramanian:Nature08}
G. Balasubramanian \textit{et al.}, %Nanoscale imaging magnetometry with
%diamond spins under ambient conditions.
\textit{Nature} \textbf{455}, 648-651 (2008).

%6
\bibitem{Maze:Nature08}
J. R. Maze \textit{et al.}, %Nanoscale magnetic sensing with an individual
%electronic spin in diamond.
\textit{Nature}, \textbf{455}, 644-647 (2008).

%7
\bibitem{Degen:APL08}
C.L. Degen %Scanning magnetic field microscope with a diamond
%single-spin sensor.
\textit{Applied Physics Letters}, \textbf{92}, 243111 (2008).

%8
\bibitem{Taylor:NatPhys08}
%J.M. Taylor  \textit{et al.},
J. M. Taylor, P. Cappellaro, L. Childress, L. Jiang, D. Budker, P.
R. Hemmer, A. Yacoby, R. Walsworth, and M. D. Lukin
%High-sensitivity diamond magnetometer with
%nanoscale resolution.
\textit{Nature Physics} \textbf{4}, 810-816 (2008).

%9
\bibitem{McGuinness:NatNano11}
L.P. McGuinness \textit{et al.}, %Quantum measurement and orientation
%tracking of fluorescent nanodiamonds inside living cells.
\textit{Nature Nanotechnology} \textbf{6}, 358-363 (2011).

%10
\bibitem{Grinolds:NatNano14}
%M.S. Grinolds \textit{et al.},
M. S. Grinolds, M. Warner, K. De Greve, Y. Dovzhenko, L. Thiel, R.
L. Walsworth, S. Hong, P. Maletinsky, and A. Yacoby \textit{Nature
Nanotechnology} \textbf{9}, 279-284 (2014).

%11
\bibitem{Tetienne:Science14}
J.-P. Tetienne \textit{et al.}, %Nanoscale imaging and control of domain-wall
%hopping with a nitrogen-vacancy center microscope
\textit{Science} \textbf{344}, 1366-1369 (2014).

%12
\bibitem{Dolde:NatPhys11}
F. Dolde \textit{et al.},
%Electric-field sensing using single diamond spins.
\textit{Nature Physics}, \textbf{7}, 459-463 (2011).

%13
\bibitem{Faklaris:Small08} O. Faklaris, D. Garrot, V. Joshi, F. Druon, J. P. Boudou,
T. Sauvage,. P. Georges, P. A. Curmi, and F. Treussart,
%Detection of single photoluminescent diamond nanoparticles in cells and study of
%the internalization pathway,
\textit{Small} \textbf{4}, 2236 (2008).

%14
\bibitem{Ermakova:NanoLett13} A. Ermakova \textit{et al.},
% G. Pramanik, J.-M. Cai, G. Algara-Siller, U. Kaiser,
%T. Weil, Y.-K. Tzeng, H. C. Chang, L. P. McGuinness, M. B. Plenio,
%B. Naydenov, and F. Jelezko, %Detection of a Few Metallo-Protein
%Molecules Using Color Centers in Nanodiamonds,
\textit{Nano Letters} \textbf{13}, 3305-3309 (2013).

%15
\bibitem{Tobar:JAP98} M.E. Tobar, J. Krupka, E.N. Ivanov, and R.A. Woode
%Anisotropic Complex Permittivity Measurements of Mono-Crystalline
%Rutile Between 10-300 Kelvin
\textit{Journal of Applied Physics} \textbf{83}, 1604 (1998).

%16
\bibitem{Krupka:MST99} J. Krupka, K. Derzakowski, M.E. Tobar, J. Hartnett, and R.G.
Geyerk, %Complex permittivity of some ultralow loss dielectric
%crystals at cryogenic temperatures,
\textit{Measurement Science and Technology} \textbf{10}, 387-392
(1999).

%17
\bibitem{LeFloch:PLA06} J.-M. Le Floch \textit{et al.},
%Whispering Modes in Anisotropic and Isotropic Dielectric Spherical Resonators,
\textit{Physics Letters A} \textbf{359}, 1-7 (2006).

%18
\bibitem{Tobar:PRD09} M.E. Tobar, E.N. Ivanov, P.L. Stanwix, J.-M. Le Floch, and J.G.
Hartnett,
% Rotating odd-parity Lorentz Invariance test in electrodynamics,
\textit{Physical Review D} \textbf{80}, 125024 (2009).

%19
\bibitem{Rieck:Proc08} C. Rieck, P. Jarlemark, R. Emardson, and K. Jaldehag, %Precision of
%time transfer using GPS carrier phase,
\textit{Proc. 22nd EFTF}, Toulouse, France, 2008.

%20
\bibitem{Droz:Proc06} F. Droz, P. Mosset, G. Barmaverain, P. Rochat, Q. Wang,
M. Belloni, L. Mattioni, F. Emma, and P. Waller,
%The On-Board Galileo Clocks - Current Status and Performance,
\textit{Proc. 20th EFTF}, Braunschweig, Germany, 2006.

%21
\bibitem{LeFloch:IEEE07} J.-M. Le Floch, M.E. Tobar, D. Cros, and J. Krupka, %High Q-factor
%Distributed Bragg Reflector Resonators with Reflectors of Arbitrary
%%Thickness,
\textit{IEEE Trans. Ultrason. Ferroelec. Freq. Contr.} \textbf{54},
2689-1695 (2007).

%22
\bibitem{LeFloch:RevSciInst14} J.-M. Le Floch \textit{et al.},
%Y. Fan, Georges Humbert, Qingxiao Shan, Denis Férachou, Romain
%Bara-Maillet, Michel Aubourg, John G. Hartnett, Valerie Madrangeas,
%Dominique Cros, Jean-Marc Blondy, Jerzy Krupka, and Michael E. Tobar
%Dielectric material characterization techniques and designs of
%high-Q resonators for applications from micro to millimeter-waves
%frequencies applicable at room and cryogenic temperatures
\textit{Review of Scientific Instruments} \textbf{85}, 031301
(2014). %doi: 10.1063/1.4867461

%23
\bibitem{LeFloch:APL08} J.-M. Le Floch, M.E. Tobar, D. Cros, and J. Krupka, %Low-loss
%Materials for high Q-factor Microwave Bragg Reflector Resonators,
\textit{Applied Physics Letters},\textbf{92}, 032901 (2008).

%24
\bibitem{Bradac:NatNano10} C. Bradac \textit{et al.}, %Observation and control of blinking
%nitrogen-vacancy centres in discrete nanodiamonds.
\textit{Nature Nanotechnology} \textbf{5}, 345-349 (2010).


%%%%%%%%%%%%%%%%%%%%%%%%%%%%%%%%%%%%%%%%%%%%%%%%%%%%%%%%%%%%%%%%%%%%%%%%%%%%%%%%%%%%%%%%%%%%%%
%%%%%%%%%%%%%%%%%%%%%%%%%%%%%%%%%%%%%%%%%%%%%%%%%%%%%%%%%%%%%%%%%%%%%%%%%%%%%%%%%%%%%%%%%%%%%%


%%
%\bibitem{Su:PRA78} C-H. Su, A.D. Greentree, W.J. Munro, K. Nemoto, and L.C. Hollenberg,
%%High-speed quantum gates with cavity quantum electrodynamics,
%\textit{Physical Review A} \textbf{78}, 062336 (2008).

%%
%\bibitem{Hong:MRS13} S. Hong, M.S. Grinolds, L.M. Pham, D. Le Sage, L. Luan, R.L.
%Walsworth, and A. Yacoby, %Nanoscale magnetometry with NV centers in
%%diamond,
%\textit{Materials Research Society} \textbf{38}, 155-161 (2013).

%%
%\bibitem{Yu:JACS05} S.-J. Yu, M.-W. Kang, H.-C. Chang, K.-M. Chen, and Y.-C. Yu,
%%Bright fluorescent nanodiamonds: no photobleaching and
%%lowcytotoxicity,
%\textit{Journal of American Chemical Society} \textbf{127}, 17604
%(2005).

%%19
%\bibitem{Soltamova:JETP2010} A.A. Soltamova \textit{et al.\ },
%\textit{JETP Letters} \textbf{92}, 102 (2010).

%%18extra
%\bibitem{Gruber:Science97}
%Gruber, A., et al., Scanning confocal optical microscopy and
%magnetic resonance on single defect centers. \textit{Science}
%\textbf{276}, 2012-2014 (1997).

%%18
%\bibitem{Oort:JPC88} E.v. Oort, N.B. Manson, and M. Glasbeek, %Optically detected spin
%%coherence of the diamond N-V centre in its triplet ground state.
%\textit{Journal of Physics C: Solid State Physics} \textbf{21}, 4385
%(1988).

%%19
%\bibitem{Amuess:PRL11} R. Ams{\"u}ss \textit{et al.\ }
%%Cavity QED with Magnetically Coupled Collective Spin States,
%\textit{Physical Review Letters} \textbf{107}, 060502 (2011).

%%20
%\bibitem{Kubo:PRB12} Y. Kubo \textit{et al.\ }
%%Electron spin resonance detected by a superconducting qubit,
%\textit{Physical Review B} \textbf{86}, 064514 (2012).

%%21
%\bibitem{Kubo:PRL10} Y. Kubo \textit{et al.\ }
%\textit{Physical Review Letters}, \textbf{105}, 140502 (2010).

%%22
%\bibitem{Schuster:PRL10} D.I. Schuster \textit{et al.\ }
%%High-Cooperativity Coupling of Electron-Spin Ensembles to
%%Superconducting Cavities
%\textit{Physical Review Letters},\textbf{105}, 140501 (2010).

%%23
%\bibitem{Healey:APL08} J.E. Healey, T. Lindström, M.S. Colclough,
%C.M. Muirhead, A.Ya. Tzalenchuk, %Magnetic field tuning of coplanar
%%waveguide resonators,
%\textit{Applied Physics Letters} \textbf{93}, 043513 (2008).

%%24
%\bibitem{Zhu:Nature11} X. Zhu \textit{et al.\ }
%%Coherent coupling of a superconducting flux qubit to an electron
%%spin ensemble in diamond
%\textit{Nature}\textbf{478}, 221-224 (2011).

%%33
%\bibitem{LeFloch:SPIE13} J-M. Le Floch, C. Bradac, T. Volz, M.E. Tobar, and S. Castelletto,
%%Non intrusive tuneable resonant microwave cavity for optically
%%detected magnetic resonance of NV-centres in nanodiamonds,
%\textit{Proc. Of SPIE - Micro+Nano Materials Devices, and Systems}
%\textbf{8923}, 89233P (2013).

%%34
%\bibitem{Zamarro:Tech87} Zamarro, %New method for the measurements of coupling
%%coefficients of transmission cavities.
%\textit{Proc.}, \textbf{134}, 103-105 (1987).

%%35
%\bibitem{Prokopenko:Tech02} Y. V. Prokopenko, and Y. F. Filippov, %Anisotropic disk
%%dielectric resonator with conducting end faces,
%\textit{Tech. Phys.} \textbf{47}, 731-736 (2002).

%%36
%\bibitem{Bradac:Nano09} C. Bradac, \textit{et al.}, %Prediction and Measurement of the
%%Size-Dependent Stability of Fluorescence in Diamond over the Entire
%%Nanoscale.
%\textit{Nano Letters}, \textbf{9},3555-3564 (2009).

%%38
%\bibitem{Loubser:RPP78} Loubser, J.H.N. and J.A. van Wyk, %Electron spin resonance in
%%the study of diamond.
%\textit{Reports on Progress in Physics} \textbf{41},1201-1248
%(1978).

%%%%%%%%%%%%%%%%%%%%%%%%%%%%%%%%%%%%%%%%%%%%%%%%%%%%%%%%%%%%%%%%%%%%%%%%%%%%%%%%%%%%%%%%%%%
%%%%%%%%%%%%%%%%%%%%%%%%%%%%%%%%%%%%%%%%%%%%%%%%%%%%%%%%%%%%%%%%%%%%%%%%%%%%%%%%%%%%%%%%%%%


\end{thebibliography}
\end{document}